\PassOptionsToPackage{unicode}{hyperref}
\PassOptionsToPackage{hyphens}{url}
\documentclass[
]{article}
\usepackage{amsmath,amssymb}
\usepackage{iftex}
\ifPDFTeX
  \usepackage[T1]{fontenc}
  \usepackage[utf8]{inputenc}
  \usepackage{textcomp} 
\else 
  \usepackage{unicode-math} 
  \defaultfontfeatures{Scale=MatchLowercase}
  \defaultfontfeatures[\rmfamily]{Ligatures=TeX,Scale=1}
\fi
\usepackage{lmodern}
\ifPDFTeX\else
\fi
\IfFileExists{upquote.sty}{\usepackage{upquote}}{}
\IfFileExists{microtype.sty}{
  \usepackage[]{microtype}
  \UseMicrotypeSet[protrusion]{basicmath} 
}{}
\makeatletter
\@ifundefined{KOMAClassName}{
  \IfFileExists{parskip.sty}{%
    \usepackage{parskip}
  }{
    \setlength{\parindent}{0pt}
    \setlength{\parskip}{6pt plus 2pt minus 1pt}}
}{
  \KOMAoptions{parskip=half}}
\makeatother
\usepackage{xcolor}
\usepackage[margin=3cm]{geometry}
\usepackage{graphicx}
\makeatletter
\def\maxwidth{\ifdim\Gin@nat@width>\linewidth\linewidth\else\Gin@nat@width\fi}
\def\maxheight{\ifdim\Gin@nat@height>\textheight\textheight\else\Gin@nat@height\fi}
\makeatother
\setkeys{Gin}{width=\maxwidth,height=\maxheight,keepaspectratio}
\makeatletter
\def\fps@figure{htbp}
\makeatother
\NewDocumentCommand\citeproctext{}{}

\makeatletter
 \let\@cite@ofmt\@firstofone
 \def\@biblabel#1{}
 \def\@cite#1#2{{#1\if@tempswa , #2\fi}}
\makeatother
\newlength{\cslhangindent}
\newlength{\csllabelwidth}
\newenvironment{CSLReferences}[2] 
 {\begin{list}{}{%
  \setlength{\itemindent}{0pt}
  \setlength{\leftmargin}{0pt}
  \setlength{\parsep}{0pt}
  \ifodd #1
   \setlength{\leftmargin}{\cslhangindent}
   \setlength{\itemindent}{-1\cslhangindent}
  \fi
  \setlength{\itemsep}{#2\baselineskip}}}
 {\end{list}}
\usepackage{calc}

\usepackage{authblk}
\usepackage{orcidlink}

\ifLuaTeX
  \usepackage{selnolig}  
\fi
\usepackage{bookmark}
\IfFileExists{xurl.sty}{\usepackage{xurl}}{} 
\hypersetup{
  pdftitle={Izzy: a high-throughput metagenomic read simulator},
  hidelinks,
  pdfcreator={LaTeX via pandoc}}

\title{Izzy: a high-throughput metagenomic read simulator}
\author[1]{Amit Lavon\orcidlink{0000-0003-3928-5907}}
\newcounter{iaffil}
\affil[\theiaffil]{Independent Researcher}
\stepcounter{iaffil}
\date{29 October 2025}

\begin{document}
\maketitle
\begin{abstract}
Simulated microbial communities are used in benchmarking microbial
abundance estimators and other bioinformatic utilities. To match current
data scales, large simulated samples are needed, and many. The speed of
current implementations might create bottlenecks for scientists testing
new innovations. Here, a new implementation is introduced, based on
existing error models. The new implementation, Izzy, provides up to a
60x speedup while maintaining a simple and easy-to-use interface.
\end{abstract}

\section{Introduction}\label{introduction}

Simulated metagenomic samples are collections of reads, meant to mimic
sequencing results of different environments. These simulated data are
helpful for developing computational methods for analyzing metagenomic
samples. Specifically, abundance estimators such as MetaPhlAn
(Blanco-Míguez et al. 2023) and Kraken (Wood, Lu, and Langmead 2019),
require samples for which underlying relative abundances are known, in
order to benchmark their estimation accuracy. This can be achieved with
sample simulation.

Among the most cited simulators is InSilicoSeq (Gourlé et al. 2019).
InSilicoSeq uses error models that were built upon real data from
several instruments. It also provides an easy-to-use user interface and
is straightforward to install, making it a popular choice. In modern
projects, however, it is normal for real samples to exceed the giga-base
scale. A sample of such magnitude may take hours to simulate, thus
simulating collections of samples could become a major bottleneck in the
development process of new tools and algorithms.

Here, a fast implementation of InSilicoSeq's error models is introduced.
The new implementation, named Izzy, aims to provide a similar user
interface while achieving higher throughput, to create an almost drop-in
replacement.

\section{Results}\label{results}

\subsection{Algorithm}\label{algorithm}

Izzy's implementation is based on InSilicoSeq's sampling algorithm, and
supports the same error models and abundance distributions. The error
models include single-nucleotide errors as well as insertions and
deletions. The abundance distributions include log-normal, half-normal,
exponential and uniform. One difference is that Izzy takes genome length
in consideration by default. Meaning that the relative share of reads
contributed by a species is proportional both to its abundance and to
its genome length. This is meant to better mimic real-world sequencing,
where larger genomes contribute more reads. This feature can be
disabled, to mimic InSilicoSeq's abundance sampling. Another feature
Izzy introduces is the ability to group together contigs of the same
genomes, using a regular-expression that is applied to each entry's
name. With this feature, each group of individual sequences that share
the same matching text with the regular-expression is considered one
species and its reads are sampled from all of its contigs.

Izzy's high speed is achieved mainly by using a statically-typed and
compiled programming language, which immediately provides a significant
speedup compared to an interpreted language such as Python. In addition,
the code is designed to eliminate dynamic (virtual) calls as much as
possible. To save storage operations, input references are read directly
without creating temporary copies, including compressed formats, and the
output files are compressed before writing, instead of writing raw files
and compressing them later.

\subsection{Performance}\label{performance}

Performance was measured by simulating reads from two datasets: the 4930
mostly-bacterial genomes from (Pasolli et al. 2019), and RefSeq's viral
reference (Pruitt, Tatusova, and Maglott 2007). Izzy and InSilicoSeq
were run on each reference, simulating samples with a hundred bacterial
species or ten viral species, log-normal distribution, and sample sizes
of 100K, 300K, 1M and 3M reads. Each experiment was repeated five times.
Izzy's throughput ranged from 1800 to 43K reads per second on the
bacterial reference, and from 29K to 173K reads per second on the viral
reference. Meanwhile, InSilicoSeq's throughput ranged from 1800 to 3200
reads per second on the bacterial reference, and stayed consistent
around 3300 on the viral reference (Figure 1). Izzy's procedure includes
a constant-time pre-processing of the input reference in each run, which
makes for a higher ratio of the run time in lower read counts. In all
tests, both programs' memory footprint did not exceed 250MB.

One caveat in the bacterial reference comparison is that Izzy was
instructed to group together contigs of the same species, while
InSilicoSeq did not have that option (it simulated individual contigs).
This was meant so simulate how Izzy would be used in a real-life
scenario, and does not affect run times.

\begin{figure}
\centering
\includegraphics[width=1\linewidth,height=\textheight,keepaspectratio]{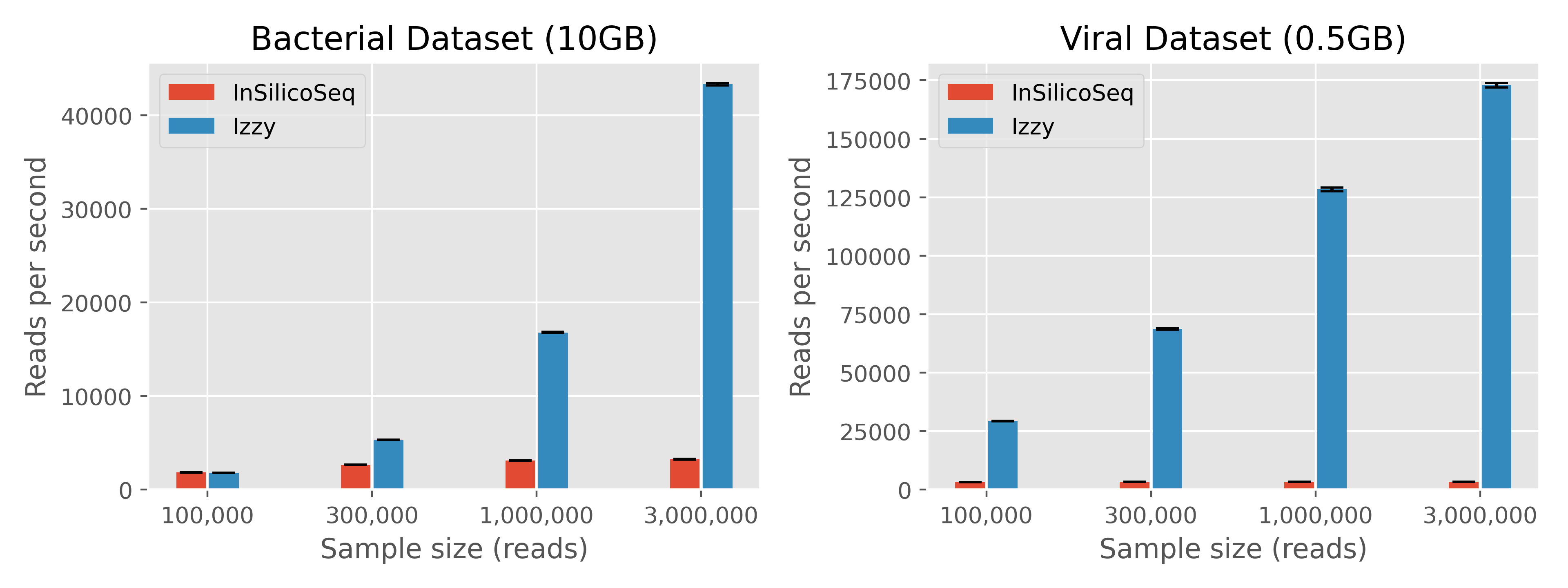}
\caption{Throughput comparison of Izzy and InSilicoSeq. Each tool was
run five times. Standard deviation is plotted in black at top of each
bar.}
\end{figure}

\section{Discussion}\label{discussion}

This project aims to bring InSilicoSeq's useful simulation model and
ease of use to large-scale projects. Izzy uses the same models and has a
similar user interface, while providing an up to 60x speedup in
throughput. It also introduces modern features, namely factoring in
genome lengths for read ratios, the ability to group together contigs of
the same species, and builtin support for compressed formats. As Izzy is
constrained to InSilicoSeq's error models, future work may include
extending the tool's support to other models, and to creating new ones
from real data. Support for multi-threading is not currently planned, as
the main bottleneck at these speeds seems to be storage throughput.
Projects that require large amounts of simulated reads, or that need
faster on-demand simulations, can save resources and wait-times by using
this implementation.

\section{Methods}\label{methods}

\subsection{Implementation}\label{implementation}

Izzy is implemented in the Go programming language, and is single
threaded. The source code is compiled into native executables for all
platforms, meaning Go is not required in order to run Izzy. The error
models are extracted from InSilicoSeq's \texttt{npz} files and are
embedded within Izzy's executable, removing dependence on external data
files.

\subsection{Benchmarking}\label{benchmarking}

InSilicoSeq version 2.0.1 was tested in this benchmark; the current
version available with \texttt{pip\ install} at the time of performing
the comparison. Performance benchmarks were run on a desktop computer
running Fedora Linux 43, with an AMD Ryzen 9 9900X CPU and 32GB RAM. The
command used for benchmarking InSilicoSeq was
\texttt{iss\ generate\ -p\ 1\ -m\ NovaSeq\ -n\ \$n\_reads\ -u\ \$n\_genomes\ -g\ \$ref\_file\ -o\ iss\ -z}.
The command for Izzy was
\texttt{izzy\ -m\ novaseq\ -n\ \$n\_reads\ -u\ \$n\_genomes\ -i\ \$ref\_file\ -o\ izz}.
Time was measured using the builtin \texttt{time} command.

\section{Availability}\label{availability}

Izzy is freely available under an MIT license as a standalone executable
at
\href{https://github.com/fluhus/izzy/releases}{github.com/fluhus/izzy/releases}
.

Feedback on the project or the manuscript is welcome at
\href{https://github.com/fluhus/izzy/discussions}{github.com/fluhus/izzy/discussions}
.

\section*{References}\label{references}
\addcontentsline{toc}{section}{References}

\protect\phantomsection\label{refs}

\end{document}